# Making BaZrS$_3$ chalcogenide perovskite thin films by molecular beam epitaxy

Ida Sadeghi[1,2], Kevin Ye[1], Michael Xu[1], James M. LeBeau[1], R. Jaramillo[1†]

1. Department of Materials Science and Engineering, Massachusetts Institute of Technology, Cambridge, MA, USA
2. Department of Electrical and Computer Engineering, University of Waterloo, Waterloo, ON, CA

† rjaramil@mit.edu

*Abstract*

We demonstrate the making of BaZrS$_3$ thin films by molecular beam epitaxy (MBE). BaZrS$_3$ forms in the orthorhombic distorted-perovskite structure with corner-sharing ZrS$_6$ octahedra. The single-step MBE process results in films smooth on the atomic scale, with near-perfect BaZrS$_3$ stoichiometry and an atomically-sharp interface with the LaAlO$_3$ substrate. The films grow epitaxially via two, competing growth modes: buffered epitaxy, with a self-assembled interface layer that relieves the epitaxial strain, and direct epitaxy, with rotated-cube-on-cube growth that accommodates the large lattice constant mismatch between the oxide and the sulfide perovskites. This work sets the stage for developing chalcogenide perovskites as a family of semiconductor alloys with properties that can be tuned with strain and composition in high-quality epitaxial thin films, as has been long-established for other systems including Si-Ge, III-Vs, and II-Vs. The methods demonstrated here also represent a revival of gas-source chalcogenide MBE.

**Introduction**

Sulfides and selenides in the perovskite and related crystal structures – chalcogenide perovskites, for brevity – may be the next family of high-performing semiconductors.[1] Chalcogenide perovskites share key physical properties with oxide perovskites, including record-high dielectric polarizability, while also featuring band gap ($E_g$) in the visible and near-infrared (VIS-NIR).[1–4] Chalcogenide perovskites are also distinguished by their good thermal and chemical stability, and non-toxic and abundant elemental components.[5,6] Chalcogenide perovskites have been demonstrated to have slow non-radiative excited-state charge recombination, direct band gap (in most cases), and strong above-band-gap optical absorption.[7–9] Theoretical predictions, preliminary experiments, and chemical intuition suggest that chemical alloying may produce materials with continuously-tunable direct band gap spanning from $E_g = 0.5$ to 2.3 eV.[4,8,10,11] These and other, related results suggest that the chalcogenide perovskite semiconductor alloy system may be useful for optoelectronic and energy-conversion technologies, particularly for solid-state lighting and solar energy conversion.

Most experimental studies to-date on chalcogenide perovskites have focused on bulk materials (*e.g.* powders) and microscopic single crystals.[11,12] Thin-film synthesis is the next, outstanding challenge. High-quality thin films are needed to enable fundamental studies of excited-state charge transport and applied studies of device performance. Thin-film synthesis may also be essential for studies of chemical doping and alloying. The good thermal stability of chalcogenide perovskites comes hand-in-hand with high materials processing temperature. The formation of compounds including low-vapor pressure refractory metals and high-vapor pressure chalcogens poses a particular challenge, common to chalcogenide perovskites and many layered and two-dimensional materials. Zr and Hf are sluggish to form crystalline compounds, requiring very high temperature for synthesis, but at very high temperatures S- and Se-containing precursors are extremely volatile, leading to chalcogen loss from the growing material, and highly-corrosive conditions for the experimental equipment. Published report of chalcogenide perovskite thin-film synthesis have appeared recently.[6,8,13–15] All reports to-date are of two-step processes that separate the processes of sulfide formation (all are sulfides to-date) and film synthesis, and all have resulted in small-grained and randomly-oriented thin films. High-quality, epitaxial, single-crystal film synthesis remains an essential goal to enable the potential of chalcogenide perovskites, as the history of other semiconductor and complex oxide materials systems teaches us.[16] Epitaxial film growth may also be able to stabilize high-selenium-content alloys in the perovskite structure, since pure selenides form instead in non-perovskite, needle-like structures.[4,17]

Here we demonstrate the making of BaZrS$_3$ thin films by molecular beam epitaxy (MBE). BaZrS$_3$ forms in the orthorhombic distorted-perovskite structure with corner-sharing ZrS$_6$ octahedra (space group Pnma, no. 62), has a direct band gap energy ($E_g$) in the range $E_g = 1.8 - 1.9$ eV, and is the most-studied chalcogenide perovskite.[9,18] The single-step MBE process results in films smooth on the atomic scale, evidenced by reflection high-energy electron diffraction (RHEED) measurements during growth, and by atomic-force microscopy (AFM) and scanning electron microscopy (SEM). The films are mirror-smooth and are brightly-colored even at 20 nm thick, indicating strong optical interaction. Epitaxial growth on LaAlO$_3$ is confirmed by X-ray diffraction (XRD) and scanning transmission electron microscopy (STEM). Films grow via two, competing epitaxial growth modes: (M1) buffered epitaxy, with a self-assembled interface layer that relieves the epitaxial strain, and (M2) direct epitaxy, with rotated-cube-on-cube growth that accommodates the large lattice constant mismatch between the oxide and the sulfide.

**Results**

We deposit films on (001)$_{PC}$-oriented LaAlO$_3$ single-crystal substrates (MTI Corp.). LaAlO$_3$ transforms from a rhombohedral (space group $R\bar{3}c$) to a cubic crystal structure at elevated temperature (approximately at 500 °C).[19] In this work, Miller indices marked with substrate "PC" (as above) indicate pseudo-cubic indexing; otherwise, we index LaAlO$_3$ in its rhombohedral structure, with room-temperature lattice constants $a = b = 5.370$ Å, $c = 13.138$ Å. For a material $ABX_3$ with corner-sharing $BX_6$ octahedra, the pseudo-cubic lattice constant $a_{PC}$ is an average $B$-$X$-$B$ distance; for LaAlO$_3$, $a_{PC} = 3.8114$ Å. Orthorhombic BaZrS$_3$ has lattice constants $a = 7.056$ Å, $b = 9.962$ Å, $c = 6.996$ Å, and $a_{PC} = 4.975$ Å.[12] The growth modes M1 and M2 are competing mechanisms to accommodate this large lattice constant mismatch.

We prepare the substrates by outgassing in the MBE chamber at 900 °C in flow of H$_2$S gas. We deposit films from elemental Ba and Zr, and H$_2$S gas. The chamber pressure during film growth varies between 5×10$^{-5}$ to 9×10$^{-5}$ torr and is controlled by the H$_2$S gas flow. For the results shown here, the H$_2$S flow rate ($Q_{H2S}$) during outgassing and film growth is between 0.6 and 0.8 sccm. The Ba and Zr rates are 0.021 Å/s and 0.0075 Å/s (approximately 1:1 stoichiometry), measured at the substrate position by a quartz crystal monitor (QCM). The substrate temperature during growth is held at 900 °C (measured at the thermocouple), and the film growth rate is 0.04 Å/s, confirmed by X-ray reflectivity (XRR). The H$_2$S gas flow is maintained during cooldown after growth. More details on the deposition methods are presented below (Methods).

Reflection high-energy electron diffraction (RHEED) data acquired during growth shows evidence of atomically-smooth, crystalline, epitaxial films. In **Fig. 1a** we present the RHEED pattern measured on the outgassed substrate along the [100]$_{PC}$ azimuth at 900 °C, before film growth. Within a short period of time after film growth starts, the substrate RHEED pattern disappears and a new pattern corresponding to the BaZrS$_3$ film appears, as shown in **Fig. 1b-c**. This pattern remains consistent throughout the whole film growth process. Quantitative analysis of the RHEED data (**Fig. 1b**) shows that the in-plane d-spacing of the film measured along the [100]$_{PC}$ substrate azimuth is 4.98 Å, corresponding (020) planes of BaZrS$_3$ (d-spacing of 4.991 Å for the relaxed structure). Measured along the [110]$_{PC}$ substrate azimuth (**Fig. 1c**), the film d-spacing is 3.52 Å, corresponding to the (121), (200), and (002) planes (d-spacings of 3.525, 3.530, and 3.513 Å, respectively for the relaxed structure). This analysis revealed that the film is fully-relaxed with the in-plane d-spacings matching their counterparts from the BaZrS$_3$ reference pattern. This is consistent with the self-assembled buffered epitaxial growth mode (M1), described below. In **Fig. 1d** we show a photograph of a typical film, which is a deep orange-reddish color, appropriate for a material with a direct band gap in the range 1.8 – 1.9 eV and strong light absorption. In **Fig. 1e** we present atomic force microscopy (AFM) data measured on a film after growth, showing that it is atomically-smooth with roughness of 3.8 Å. In **Fig. 1f** we present the result of the XRR measurement on the film. The presence of well-defined Kiessig interference fringes in the XRR curve indicates that the film surface is smooth and the film-substrate interface is well defined; by modeling the XRR data we find the film thickness to be 23.52 ± 0.21 nm.

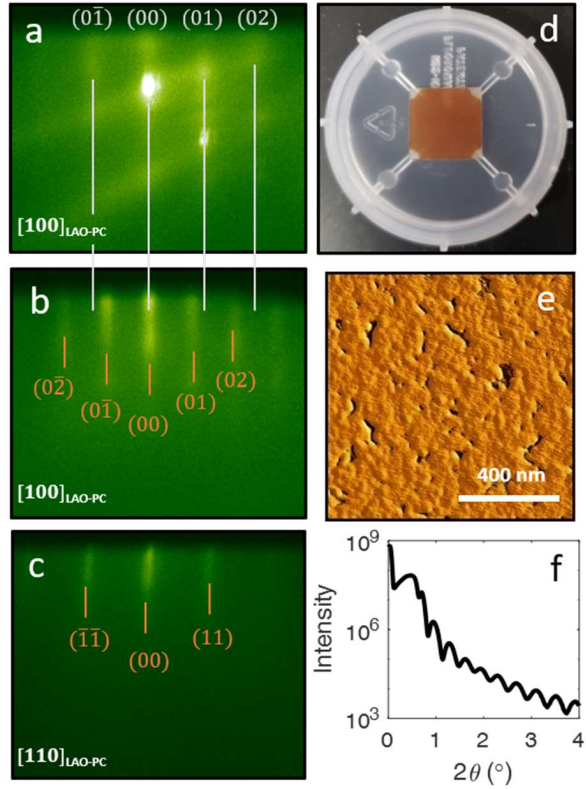

**Figure 1:** Growth of smooth and epitaxial BaZrS$_3$ thin films on LaAlO$_3$ by MBE. (a) RHEED data measured on a LaAlO$_3$ substrate at 900 °C before the start of film growth. The indices and light grey lines mark reflections measured along the substrate [100]$_{PC}$ azimuth. (b, c) RHEED data measured during film growth with the sample oriented along the [100]$_{PC}$ and [110]$_{PC}$ substrate azimuths, respectively. The RHEED data in (a-c) are scaled and centered similarly, so that the spacing of the different reflections can be compared directly. The substrate (light grey) and film (orange) reflections are labeled using pseudo-cubic indexing. (c) Photograph of a typical sample; film is 23.7 nm thick and is uniform across the 1 cm$^2$ substrate. (d) AFM data showing a smooth surface interrupted by depressions; the image root-mean-square roughness is 3.92 Å. (f) XRR result showing Kiessig fringes indicating a smooth surface.

In **Fig. 2** we present x-ray diffraction (XRD) data. The out-of-plane scan (**Fig. 2a**) shows the BaZrS$_3$ (H0K) family appearing with the LaAlO$_3$ (00L) family of reflections. The rocking curve of the (202) film peak has a full-width of 0.409°. Two competing epitaxial growth modes contribute to this rocking curve width, as discussed below. In **Fig. 2b** we present azimuthal ($\varphi$) scans that identify the in-plane epitaxial relationship between film and substrate. The film (200), (121) and (002) peaks align with the substrate (110) peaks, corresponding to alignment of the pseudo-cubic edges of the film and substrate crystal structures (M1). We also observe a secondary, weaker set of film (200), (121) and (002) reflections that are offset by 45° from the primary reflections and that correspond to rotated-cube-on-cube growth (M2).

We use the XRD data to analyze the film strain state. We measure the lattice parameter using coupled $\theta/2\theta$ scans of the film peaks (202), (200), (121) and (002), focus on the dominant set of reflections found in the phi scan, and we model the film as a material with tetragonal symmetry (with lattice constants $a = c$) to simplify the analysis. We find $a = c = 7.0624 \pm 0.0055$ Å and $b = 9.9923 \pm 0.0117$ Å, which agree well with reference room-temperature lattice constants (above). We conclude that the dominant growth mode (M1) with aligned pseudo-cubic edges results in a fully-relaxed film.

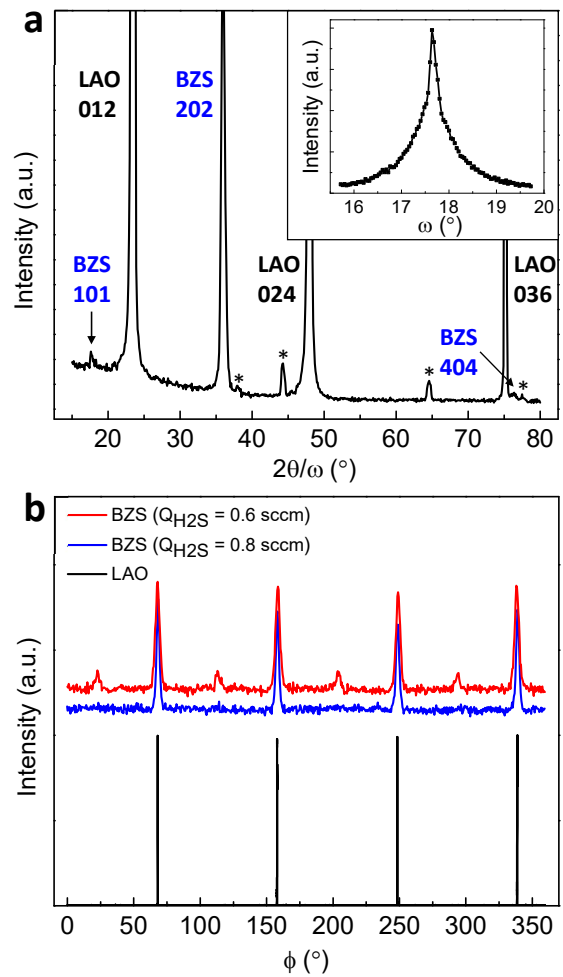

**Figure 2:** X-ray diffraction data for BaZrS$_3$ (BZS) films grown on LaAlO$_3$ (LAO). (a) Out-of-plane $2\theta/\theta$ XRD scan, sample grown with $Q_{H2S} = 0.8$ sccm. Inset shows the rocking curve of the BaZrS$_3$ (202) reflection. Peaks marked with asterisks are instrument background signals. (b) $\phi$ scan showing that the film (200), (121), and (002) peaks align with the substrate (110) peaks (growth mode M1). We also observe a weaker set of film

reflections rotated by 45° from the substrate (110) peaks (growth mode M2) for the film growth with lower $Q_{H2S}$.

To better understand the film structure and epitaxial growth modes, we turn to scanning transmission electron microscopy (STEM). From the high angle annular dark-field (HAADF) STEM overview, **Fig. 3a**, we see that the film thickness is largely uniform, 23 ± 0.5 nm. Distributed along the film are notched locations, approximately 20 - 40 nm wide, where the film thickness decreases to 17.9 ± 1.2 nm, marked with red arrows in **Fig. 3a**. At the atomic scale, HAADF STEM data confirms that the sulfide $BaZrS_3$ thin film grows epitaxially on the oxide $LaAlO_3$ substrate (**Fig. 3b-c**). The predominate epitaxial relationship is $(101)_{BZS} \parallel (001)_{LAO-PC}$ and $[010]_{BZS} \parallel [010]_{LAO-PC}$, meaning that the pseudo-cubic edges of the substrate and film are parallel (growth mode M1) (**Fig. 3b**).. At the interface, the structure differs from the bulk of the thin film. The atomic number-sensitive HAADF intensity is significantly decreased over two atomic planes, likely due to compositional differences and/or static displacements along the column.[20,21] Further, the interface is incommensurate with the substrate, with the coincident site lattice (CSL) alignment roughly every five substrate planes and four film planes. The La atom column intensity drops at the CSL positions, suggesting vacancies or off-stoichiometry. The film appears to be completely relaxed with no measurable epitaxial strain within and past the "buffer" layer.

At the notched locations marked in **Fig. 3a**, there are substrate step edges, and the epitaxial relationship changes to $(101)_{BZS} \parallel (001)_{LAO-PC}$ and $[11\bar{1}]_{BZS} \parallel [010]_{LAO-PC}$ (**Fig. 3c**). This is the rotated-cube-on-cube relationship (growth mode M2). These regions represent approximately 20 % of the observed area and produce the weak 45°-offset reflections seen in the XRD ɸ-scan (**Fig. 2b**). In these regions, there is a significant amount of plane bending that accommodates the step edge and the surrounding M1-oriented regions (**Fig. 3d**). This inclination may be due to nucleation and growth on substrate step edges, contributing to the width of the out-of-plane reflection rocking curves (**Fig. 2a**). At the interface, the HAADF contrast is significantly lower than the film, with film and substrate appearing to be directly bonded due to local atom column distortions at the interface, visible in the Fourier-filtered image (**Fig. 3e**). An extra plane in the film, or misfit dislocation, is present roughly every 14 planes in the substrate. Comparing the interplanar distances at this orientation, $d_{BZS} = 3.5213$ Å and $d_{LAO} = 3.8114$ Å, we conclude that the misfit dislocations accommodate a majority of the lattice constant mismatch, lowering the film strain from 8% to 1%. Away from the interface, there is no visible strain gradient, with projected interatomic spacings in agreement with the bulk crystal structure. In these regions of growth mode M2, the film thickness is reduced, indicating a slower growth rate than for growth mode M1. This is likely due to the higher strain energy associated with the directly-bonded interface, resulting in thermodynamically less-favored growth compared to the immediately-relaxed "buffered" growth seen in the M1 structure, where evidence for direct bonding is absent.

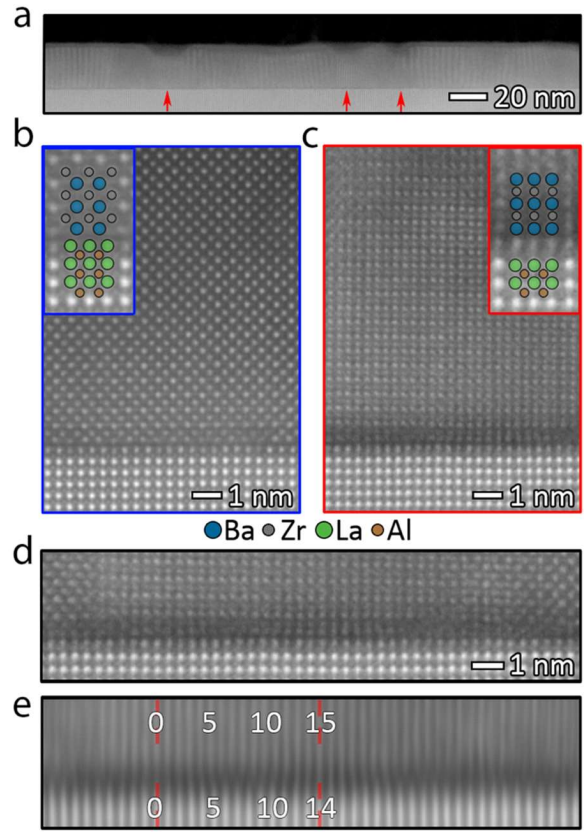

**Figure 3:** HAADF STEM data for a BaZrS$_3$ film grown on an LaAlO$_3$ substrate. (a) Overview image showing near-uniform thickness interrupted by "notch" features (marked with red arrows). (b) Atomic-resolution images showing two epitaxial growth modes: the predominate mode with pseudocubic edges aligned (growth mode M1, left panels with blue boxes), and (c) the rotated-cube-on-cube mode with direct bonding (growth mode M2, right panels with red boxes). The insets are magnified images with overlaid atomic species. (d) Larger-area image showing coincidence of a substrate step edge and a region of growth mode M2, flanked by regions of growth mode M1. (e) The Fourier-filtered image of (d), which highlights the misfit dislocation that accommodates 15 film lattice planes for every 14 substrate lattice planes.

From energy dispersive X-ray spectroscopy (EDS) in STEM, we find that the composition of the "bulk" film is nearly stoichiometric BaZrS$_3$: (in at%) Ba: 19.25 +/- 2.69, Zr: 20.60 +/- 3.20, S: 60.15 +/- 5.39. Atomic-resolution EDS maps show the perovskite structure of the BaZrS$_3$ films (**Fig. 4a**). To determine chemical interdiffusion, we use electron energy loss spectroscopy (EELS) due to its greater sensitivity to oxygen detection, relative to EDS. The EELS maps reveal that that there is negligible interdiffusion across the interface, with coincidence of film and substrate species limited to a 1 nm-wide region (**Fig. 4b**). Importantly, we see from the line profiles (**Fig. 4c**) that the oxygen signal follows the same trend as the La and Al signals, and that all are undetectable within the "bulk" of the film, away from the narrow interface region. Similarly, the integrated Ba,

Zr, and S signals become undetectable in the substrate. We do observe an increase in oxygen signal at the film top surface. This topmost 5 nm of the film appears to be non-stoichiometric $BaZrS_3$, likely a result of exposure to oxygen or sample preparation (**Fig. 4b-c**). We hypothesize that sulfur loss from the near surface region during sample processing combined with the thermodynamic driving force for Zr metal oxidation may lead to this non-stoichiometry.

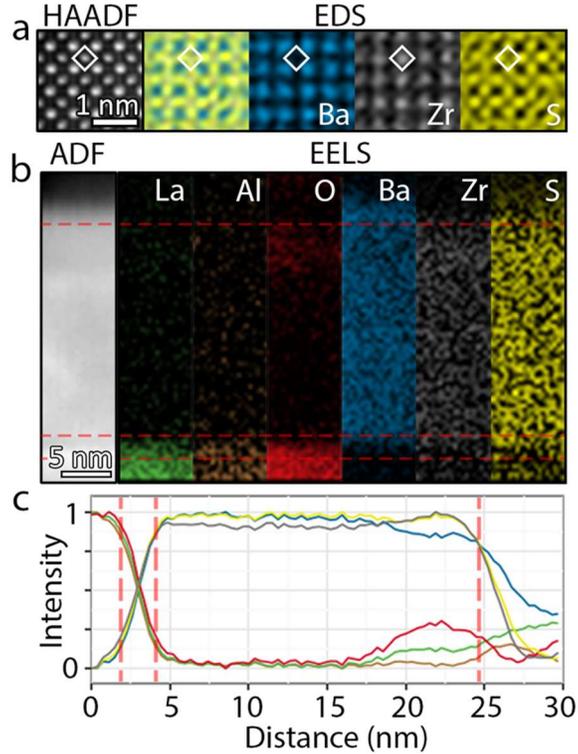

**Figure 4:** EDS and EELS maps and line profiles showing little interdiffusion at the film-substrate interface, a uniform "bulk" $BaZrS_3$ film, and partial oxidation at the top surface. (a) HAADF STEM image of the film and corresponding atomic resolution EDS maps of Ba, Zr, and S, processed using a radial Wiener filter (b) Annular dark-field overview of the spectrum image area and EELS intensity maps derived from the La-M, Al-K, O-K, Ba-M, Zr-L, and S-K edges. (c) Integrated and denoised intensity profiles extracted from the data in (b).

We find planar defects throughout the film, corresponding to boundaries between rotation variants, as well as antiphase boundaries. In **Fig. 5a** we highlight a boundary between 90° variants of the M1 growth mode. We find that the $\hat{b}$ direction remains in-plane for all M1 domains. However, given the four-fold symmetry of the substrate, the in-plane orientations $[10\bar{1}]_{BZS} \parallel [010]_{LAO\text{-}PC}$, $[010]_{BZS} \parallel [010]_{LAO\text{-}PC}$, and $[\bar{1}01]_{BZS} \parallel [010]_{LAO\text{-}PC}$ are equally favorable, creating planar boundaries where these domains meet. These boundaries introduce local changes in the $ZrS_6$ octahedra tilt and distortion patterns that are likely correlated with the band-edge electronic structure and semiconducting properties (*e.g.* band gap).[18] At substrate step edges we find antiphase boundaries in the film, as shown in **Fig. 5b**. The normalized intensities of atom columns near the planar defect emphasize differences in Z-contrast and highlight the antiphase boundary. Atom columns are also visible in positions associated with the rotated-cube-on-cube. Atom columns corresponding to growth mode M2, rotated-cube-on-cube, are also faintly visible. This is consistent with the hypothesis that M2 growth nucleates at substrate steps.

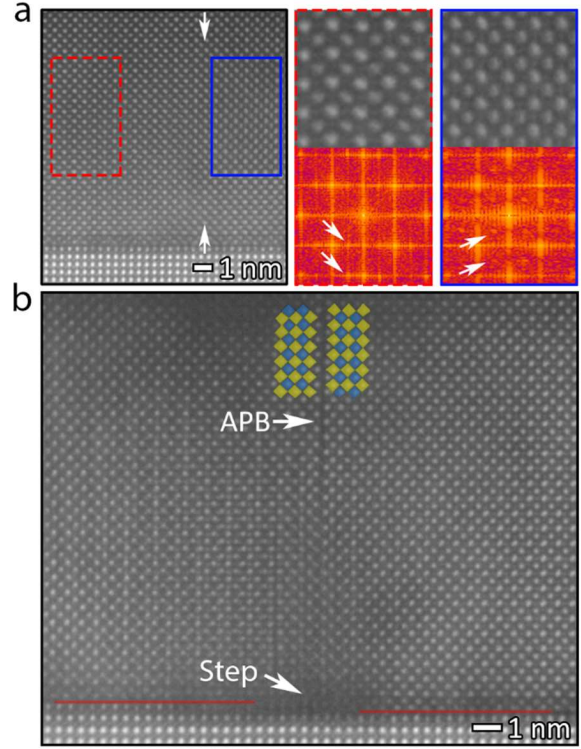

**Figure 5:** STEM HAADF images showing planar defects and rotation variants in a BaZrS$_3$ film grown on an LaAlO$_3$ substrate. (a) Boundary between two 90° rotation variants of growth mode M1 occur throughout the film. Enlarged images and fast-Fourier transforms show the difference between the two rotation variants in (a), with unique superlattice reflections marked. (b) Antiphase boundaries are located at substrate step edges. The yellow and blue squares denote the dimmer (Zr/S) and brighter (Ba) atom columns, respectively, adjacent to the antiphase boundary. Atom columns corresponding to growth mode M2 are also faintly visible in the left-center half of the image, which may nucleate at substrate steps.

The combined XRD and STEM data allow us to describe the two epitaxial growth modes, in context with the challenges present in epitaxial chalcogenide perovskite thin film growth. Chalcogenide perovskites including BaZrS$_3$ have much larger unit cells than their oxide counterparts. This poses a challenge to identify appropriate substrates for cube-on-cube epitaxy. In **Fig. 6a** we compare $a_{PC}$ of select chalcogenide perovskites to those of some commonly-available substrate materials. We present the sulfide perovskites BaZrS$_3$ and SrHfS$_3$; we also present the selenide BaZrSe$_3$, which potentially could be stabilized in the perovskite structure through epitaxy.[4,17] Chalcogenide perovskites feature $a_{PC} \gtrsim 5$ Å, whereas typical perovskite oxides feature $a_{PC} \approx 3.7 - 4.1$ Å. This large lattice constant mismatch can be accommodated by a 45° rotation between film and substrate, referred to as rotated-cube-on-cube growth (or "root-2 epi"). We label this as direct epitaxy, growth mode M2. For BaZrS$_3$ ($a_{PC}$ = 4.975 Å) on LaAlO$_3$ ($a_{PC}$ = 3.8114 Å), this results in a tensile strain of 8%. STEM data shows that a majority of this strain is relieved by the presence of misfit dislocations, with residual strain visible in the rotated direct bonds at the interface (**Fig. 3**). We also observe that this material grows more slowly than the rest of the film, resulting in a notched film surface. Meanwhile, the majority of the film grows in a relaxed state on a self-assembled interface ("buffer") layer that fully relieves the epitaxial strain. The pseudo-cubic edges of film and substrate align, but there is no direct bonding between the perovskite structure of film and substrate. We label this as buffered epitaxy, growth mode M1. M1 growth is faster than M2, and presumably dominates everywhere except for where M2 domains have nucleated at step edges and grown to a critical size. The balance between M1 and M2 changes with the H$_2$S flow rate. Evidence for M1 growth disappears from the XRD data when $Q_{H2S}$ changes from 0.6 to 0.8 sccm.

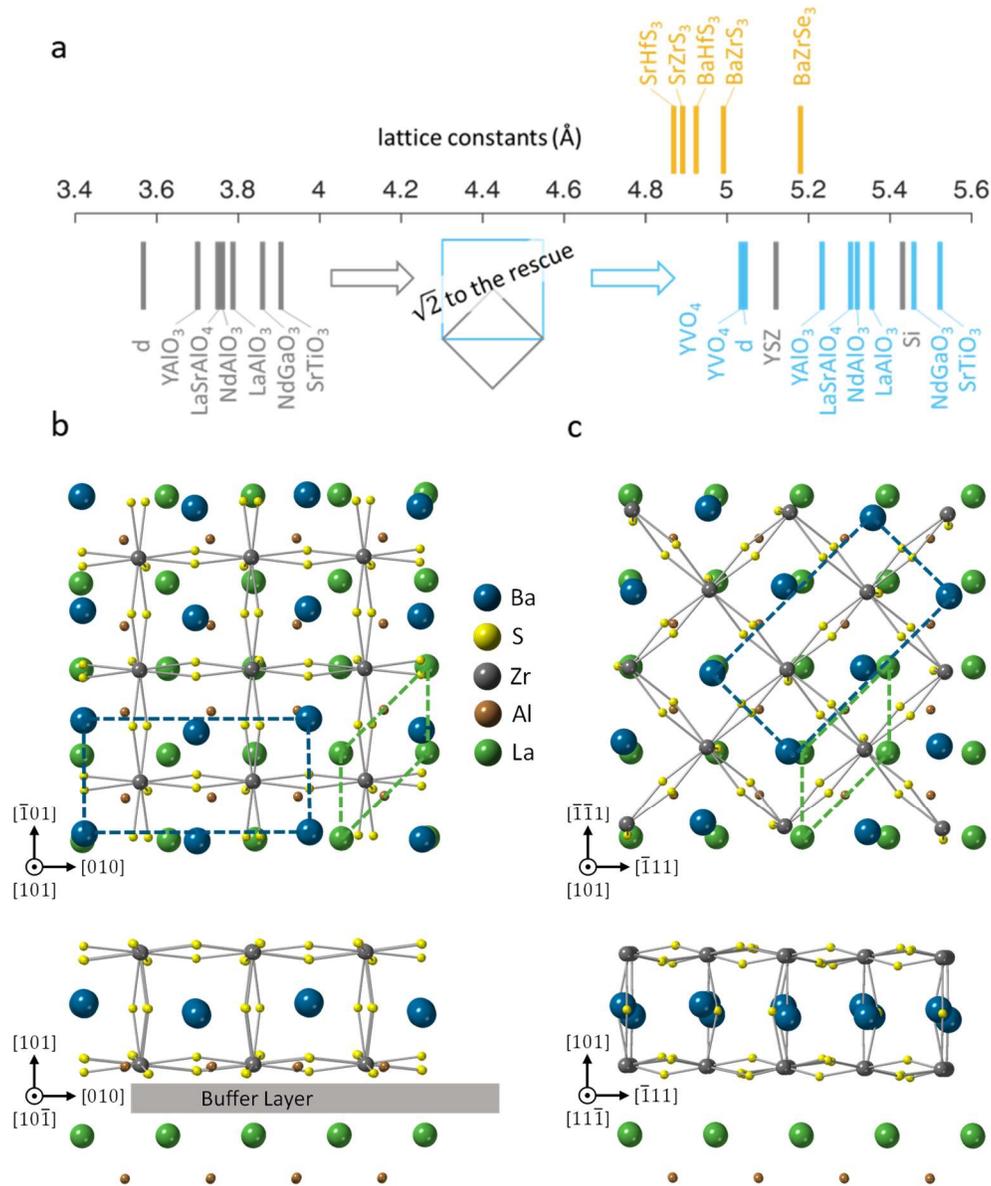

**Figure 6:** Epitaxial growth modes for chalcogenide perovskite films on oxide perovskite substrates. (a) Comparing the pseudo-cubic lattice constants of chalcogenide perovskites and commercially-available crystal substrates; d = diamond, YSZ =yttria-stabilized zirconia. Orange = select chalcogenide perovskites. Grey = pseudo-cubic lattice constants of commercially-available substrates. Blue = select pseudo-cubic lattice constants scaled by $\sqrt{2}$. (b) Epitaxial relationship for $BaZrS_3$-on-$LaAlO_3$ growth mode M1, buffered epitaxy. (c) Epitaxial relationship for $BaZrS_3$-on-$LaAlO_3$ growth mode M2, direct epitaxy. In (b) and (c), the crystal structures are viewed in orthographic projection in their unstrained state, and oxygen is omitted from the substrate for clarity. The top row shows

plan-views, with the unit cells illustrated by dashed lines (blue rectangle for BaZrS$_3$, green parallelogram for LaAlO$_3$). The bottom row shows side-views.

**Conclusion**

This work sets the stage for developing chalcogenide perovskites as a family of semiconductor alloys with properties that can be tuned with strain and composition in high-quality epitaxial thin films, as has been long-established for other systems including Si-Ge, III-Vs, and II-Vs. This work also relates to the established field of complex oxide epitaxial growth and materials physics. Our work suggests that extended defects and the tilt and distortions of ZrS$_6$ octahedra may be controlled through epitaxy, as they are in complex oxides. In the oxides, these structural parameters couple to properties such as dielectric response, magnetism, and insulator-metal transitions, via strong correlations between atomic and electronic structure. Chalcogenide perovskites exhibit similar correlations, with the notable difference that, as semiconductors with band gap in the VIS-NIR spectral region, we expect to control properties useful for optoelectronics such as band gap, luminescence yield, and charge transport mobility. Our work sets the stage for controlling such properties through strain, doping, and alloying, and for device demonstrations and applications. The methods demonstrated here also represent a revival of gas-source chalcogenide MBE, with tremendous potential for impact on research on chalcogenide perovskites and other sulfur- and selenium-containing compounds.

**Methods**

We grow films in a chamber custom-modified for gas-source chalcogenide MBE (Mantis Deposition M500). The substrate is heated radiatively from a SiC filament, and is rotated at 2 rpm. Ba metal is supplied from an effusion cell (Mantis Comcell 16-500), and Zr metal from an electron-beam (e-beam) evaporator (Telemark model 578). We use ferrite clamps on power cables running between the control rack and the MBE chamber to reduce the impact of e-beam arcing events on other components; left unaddressed, e-beam arcs can cause other components including effusion cells and the substrate heater to shut down. Sulfur is supplied in the form of H$_2$S gas. The gas is supplied from a condensed, liquified source of 99.9% purity (Matheson Research Grade), and runs through a purifier (Matheson Purifilter) before entering the MBE chamber. The gas is injected in close proximity to the substrate using custom-made gas lines and nozzles (SS310 construction), so that the pressure experienced at the substrate is higher than the chamber pressure reading. We estimate from Monte-Carlo gas flow simulations (using MolFlow+, results not shown) that the H$_2$S pressure at the substrate during growth is 50% higher than the chamber pressure.

RHEED data are measured using a 20 keV electron gun (Staib) and a digital acquisition system (k-Space Associates, kSA 400). The strong magnetic field from the permanent magnets in the electron-beam evaporator interferes with RHEED measurements. The magnetic field strength reaches as high as ~10 Oe along the RHEED electron beam path; such a strong field distorts the RHEED pattern, and can even cause the beam to miss the imaging screen entirely. We partially address this problem by installing custom magnetic shielding along much of incident and reflected beam paths, so that the RHEED beam path is unshielded for only several inches, as it encounters the substrate. Even so, the RHEED beam cannot be centered on the screen, and the pattern at times appears tilted. This does not affect the analysis presented here.

We perform XRR measurements using a Rigaku Smartlab in parallel-beam mode, with a Cu target, and a tube power of 9 kW (45 kV, 200 mA). We perform XRD using a Bruker D8 High-Resolution X-ray diffractometer with a Ge (022) four-bounce monochromator in parallel-beam mode, with a Cu target, and a tube power of 1.6 kW (40 kV, 40 mA).

Samples for STEM measurements are prepared by non-aqueous wedge polishing, followed by single-sector ion milling (Fischione 1051) using ion-beam energies of 2 kV, 1 kV, and 0.5 kV.[22,23] STEM data is acquired on a probe-aberration-corrected Thermo Fisher Scientific ThemisZ S/TEM operated at 200 kV using a convergence angle of 17.9 mrad. HAADF images are acquired using a collection angle of 63-200 mrad, and drift-corrected using the Revolving STEM (RevSTEM) method.[24,25] EDS is performed at low magnification using Super X detectors at a beam current of 500 pA and quantified using the Thermo Fisher Scientific Velox software. Atomic-resolution EDS maps are filtered using a radial Wiener filter with a high frequency cutoff of 0.08 px$^{-1}$. Electron energy loss spectroscopy is performed using a Gatan Continuum spectrometer with an energy dispersion of 1.5 eV/ch, collection angle of 100 mrad, and a sample thickness of approximately 60 nm.[26] The resulting spectra are extracted using the HyperSpy python package and denoised using principle component analysis (PCA), retaining only the first four components, which accounts for 96.8% of the variance.[27]

**Acknowledgments**

We acknowledge support from the National Science Foundation (NSF) under grant no. 1751736, "CAREER: Fundamentals of Complex Chalcogenide Electronic Materials," and from the Office of Naval Research under grant no. N00014-18-1-2746. A portion of this project was funded by the Skolkovo Institute of Science and Technology as part of the MIT-Skoltech Next Generation Program. KY acknowledges support by the NSF Graduate Research Fellowship, grant no. 1745302. JML and MX acknowledge support from the Air Force Office of Scientific Research (FA9550-20-0066) and the MIT Research Support Committee. This work was carried out in part through the use of the MIT Materials Research Laboratory (MIT MRL) and MIT.nano facilities. We acknowledge helpful discussions with Jayakanth Ravichandran of the University of Southern California, whose research group recently achieved similar chalcogenide epitaxial film growth using pulsed laser deposition.